\documentclass[fleqn,twoside]{article}
\usepackage{espcrc2}

\usepackage{graphicx}
\usepackage[figuresright]{rotating}

\newcommand{\AmS}{{\protect\the\textfont2
  A\kern-.1667em\lower.5ex\hbox{M}\kern-.125emS}}

\title{Extracting the fundamental parameters}

\author{Dirk Zerwas\address[LAL]{LAL, 
        CNRS/IN2P3, 
        B.P.34, 91898 Orsay Cedex, France}}
       
\begin{document}

\begin{abstract}
If supersymmetry is discovered at the LHC, the extraction of the fundamental parameters will be a 
formidable task. In such a system where measurements depend on different combinations of the 
parameters in a highly correlated system, the identification of the true parameter set in an 
efficient way necessitates the development and use of sophisticated methods. A rigorous treatment 
of experimental and theoretical errors is necessary to determine the precision of the measurement 
of the fundamental parameters. The techniques developed for this endeavor can also be applied to 
similar problems such as the determination of the Higgs boson couplings at the LHC. 
\vspace{1pc}
\end{abstract}

\maketitle

\section{INTRODUCTION}

In the following two scenarios for the extraction of the 
fundamental parameters from experimental data will be discussed. 
The first example is vintage supersymmetry as an example 
for the precision achievable at the LHC when supersymmetry provides a multitude
of signatures. The second example is difficult supersymmetry, be it that 
most, but not all of the supersymmetric partners are not observed, or  
that only a light Higgs boson is observed and its properties
measured at the LHC.

In a supersymmetric theory, each fermionic degree of freedom has a bosonic 
counter part and vice versa. Squarks, sleptons, neutralinos and charginos
are the partners of the quarks, leptons, neutral and charged gauge and 
Higgs bosons.
Supersymmetric theories have ``no'' problems
with radiative corrections, they predict a light Higgs boson with a mass 
less than 150~GeV, provide interesting phenomenology at the 
the TeV scale and may provide a link to the Planck scale.

Many different models describe the phenomenology of supersymmetry. In the following
sections, three of these will be discussed: mSUGRA as an example of a model 
with few parameters, most of which are defined at the grand unification scale (GUT),
Decoupled Scalars Supersymmetry (DSS) as an example where a part of the supersymmetric
spectrum is unobservable at the LHC, and the MSSM as a
an example of a model defined at the electroweak scale with many parameters. 
Determining this model (in contrast to mSUGRA) may allow to measure grand unification
of the supersymmetric breaking parameters. 

R--parity is conserved in the following thus supersymmetric
particles (cascade-)decay to the lightest supersymmetric particle (LSP). In the
following the lightest neutralino will play the role of the LSP. It is stable, 
neutral and weakly interacting. The LSP is a candidate for 
dark matter. The experimental signature for supersymmetry
at colliders is missing transverse energy due to the presence of the undetected 
LSP(s).

In the present absence of supersymmetric signals it is necessary to define benchmark 
points in order to study the potential of supersymmetry discovery and measurement. 
The reference point SPS1a~\cite{Allanach:2002nj} has been studied intensely in the past 
years in LHC as well as ILC simulations~\cite{Weiglein:2004hn}:
In SPS1a gluinos and squarks have masses of the order of 500--600~GeV,
light sleptons have masses of 150~GeV and the Higgs boson is at the LEP limit.

\section{DETERMINATION OF PARAMETERS}

The difficulties of determining the fundamental parameters from measurements can be illustrated
by taking mSUGRA as an example. The mass of the smuon depends on m$_0$, m$_{1/2}$ and 
$\tan\beta$. The mass of the chargino depends on m$_{1/2}$ and $\tan\beta$, thus the 
same parameter has an impact on different measurements. Additionally the experimental 
errors can be correlated and each theoretical prediction also has an error associated
to it. Thus in order to disentangle the system to obtain the best possible 
precision on all parameters, a global Ansatz is necessary. 

For each observable in addition to a precise ``experimental'' determination, a precise
theory prediction must be associated. This means the most 
up to date calculations for masses, branching ratios, cross sections as well
as dark matter predictions have to be used. A summary of different tools 
is listed in~\cite{Allanach:2008zn}.  

The first study of this type was performed in~\cite{Blair:2002pg}. The 
Fittino~\cite{Bechtle:2005vt} and SFitter~\cite{Lafaye:2004cn,Lafaye:2007vs}
collaborations studied the reconstruction of parameters 
with sophisticated techniques which will be described later.

\section{PREDICTIONS FROM PRESENT DATA}

While no direct observation of a supersymmetric particle is available, the theoretical
predictions of precision observables are sensitive, via radiative corrections,
to the parameters of supersymmetry. A wealth of precise measurements such as the mass of the 
W~boson and the top quark mass have been made at LEP~\cite{Alcaraz:2007ri}
and the TeVatron~\cite{Varnes:2008tc}. 
Additionally precise measurements in the b--sector, such as the branching ratio
of b$\rightarrow$s$\gamma$ have been performed. The anomalous
magnetic moment of the muon ($(g-2)_\mu$) is known with very good precision and 
WMAP has provided a measurement of the relic density. 

\begin{figure}[htb]
\includegraphics[width=\columnwidth]{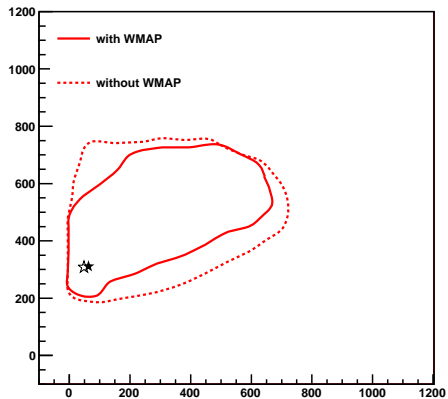}
\vspace{-1.5cm}
\caption{The best fit result is shown in the $(\mathrm{m}_{1/2}/\mathrm{m}_0)$ plane
as well as the contours including and excluding the WMAP measurements.}
\label{fig:EWPOsusy}
\end{figure}

The comparison
of the predictions with the measurements allows to determine the parameters 
of mSUGRA~\cite{Buchmueller:2008qe}. Figure~\ref{fig:EWPOsusy} shows the best--fit point in the 
$(\mathrm{m}_{1/2}/\mathrm{m}_0)$ plane. Due to $(g-2)_\mu$ a positive sign of $\mu$
is favored.
It is intriguing to note that the SPS1a point,
a point considered to be overly optimistic at the time of its definition, is 
close to the best--fit point. 

This approach can be taken a step further by fitting the phenomenological
MSSM, a weak scale model with about 20 supersymmetric parameters.
Using a Bayesian approach with a linear prior and a logarithmic prior, one 
can infer on the MSSM parameters as shown in~\cite{AbdusSalam:2009qd}.
The dependence of the results on the prior should disappear 
as more information (such as supersymmetric mass measurements) become available.
It is interesting to note that in this study the preference for a positive
sign of the $\mu$ parameter from the muon anomalous magnetic moment is 
not observed. 

\section{VINTAGE SUPERSYMMETRY}

\begin{table*}[htb] 
\caption[]{ 
LHC measurements~\cite{Weiglein:2004hn} in SPS1a are shown. The nominal values are calculated with SuSpect.
  Statistical errors, systematic errors from the lepton (LES), 
  jet energy scale (JES) and theoretical errors are all given in GeV.} 
\begin{small} \begin{center} 
\begin{tabular}{|ll|r|rrrr|} 
\hline 
\multicolumn{2}{|c|}{ type of } &  
 \multicolumn{1}{c|}{ nominal } &  
 \multicolumn{1}{c|}{ stat. } &  
 \multicolumn{1}{c|}{ LES } &  
 \multicolumn{1}{c|}{ JES } &  
 \multicolumn{1}{c|}{ theo. } \\ 
\multicolumn{2}{|c|}{ measurement } &  
 \multicolumn{1}{c|}{ value } &  
 \multicolumn{4}{c|}{ error } \\ 
\hline 
\hline 
$m_h$ &  
 & 108.99& 0.01 & 0.25 &      & 2.0 \\ 
$m_t$ &  
 & 171.40& 0.01 &      & 1.0  &     \\ 
$m_{\tilde{l}_L}-m_{\chi_1^0}$ &  
 & 102.45& 2.3  & 0.1  &      & 2.2 \\ 
$m_{\tilde{g}}-m_{\chi_1^0}$ &  
 & 511.57& 2.3  &      & 6.0  & 18.3 \\ 
$m_{\tilde{q}_R}-m_{\chi_1^0}$ &  
 & 446.62& 10.0 &      & 4.3  & 16.3 \\ 
$m_{\tilde{g}}-m_{\tilde{b}_1}$ &  
 & 88.94 & 1.5  &      & 1.0  & 24.0 \\ 
$m_{\tilde{g}}-m_{\tilde{b}_2}$ &  
 & 62.96 & 2.5  &      & 0.7  & 24.5 \\ 
$m_{ll}^\mathrm{max}$: & three-particle edge($\chi_2^0$,$\tilde{l}_R$,$\chi_1^0$)   
 & 80.94 & 0.042& 0.08 &      & 2.4 \\ 
$m_{llq}^\mathrm{max}$: & three-particle edge($\tilde{q}_L$,$\chi_2^0$,$\chi_1^0$)   
 & 449.32& 1.4  &      & 4.3  & 15.2 \\ 
$m_{lq}^\mathrm{low}$: & three-particle edge($\tilde{q}_L$,$\chi_2^0$,$\tilde{l}_R$) 
 & 326.72& 1.3  &      & 3.0  & 13.2 \\ 
$m_{ll}^\mathrm{max}(\chi_4^0)$: & three-particle edge($\chi_4^0$,$\tilde{l}_R$,$\chi_1^0$) 
 & 254.29& 3.3  & 0.3  &      & 4.1 \\ 
$m_{\tau\tau}^\mathrm{max}$: & three-particle edge($\chi_2^0$,$\tilde{\tau}_1$,$\chi_1^0$) 
 & 83.27 & 5.0  &      & 0.8  & 2.1 \\ 
$m_{lq}^\mathrm{high}$: & four-particle edge($\tilde{q}_L$,$\chi_2^0$,$\tilde{l}_R$,$\chi_1^0$) 
 & 390.28& 1.4  &      & 3.8  & 13.9 \\ 
$m_{llq}^\mathrm{thres}$: & threshold($\tilde{q}_L$,$\chi_2^0$,$\tilde{l}_R$,$\chi_1^0$) 
 & 216.22& 2.3  &      & 2.0  & 8.7 \\ 
$m_{llb}^\mathrm{thres}$: & threshold($\tilde{b}_1$,$\chi_2^0$,$\tilde{l}_R$,$\chi_1^0$) 
 & 198.63& 5.1  &      & 1.8  & 8.0 \\ 
\hline 
\end{tabular} 
\end{center} \end{small} \vspace*{-3mm} 
\label{tab:edges} 
\end{table*} 

The point SPS1a provides a vast number of measurements at the LHC.
In particular, the long cascade decay 
$\tilde{q}_L\rightarrow\chi_2^0 q\rightarrow\tilde{\ell}_R\ell q\rightarrow\ell\ell q \chi^0_1$
can be observed well above the Standard Model and supersymmetric backgrounds. 
The final state consists of opposite-sign same flavor leptons, i.e., electrons
and muons, and hard jets.
In this decay chain edges and thresholds can be measured by reconstructing 
invariant masses of different combinations: lepton--lepton, lepton--jet, 
lepton--lepton--jet. These endpoints are analytical functions of the masses
of the particles, they do not depend on the underlying theoretical model.

The results of the LHC study assuming an integrated luminosity of 300~fb$^{-1}$
are shown in Table~\ref{tab:edges}. Several of the measurements 
are limited by the systematic error on the energy scale, which is assumed
to be of the order of percent for jets and at the per mil level for leptons.
Using the kinematic formula for the edges and endpoints, the absolute masses
of the supersymmetric particles can be determined. This procedure, usually performed
with fits on toy experiments introduces additional correlations. For the determination
of the fundamental parameters it is therefore preferable to start directly from the
edges and thresholds. Simplifying the results of the analysis for the ILC, 
as soon as the particles are kinematically accessible, they can
be measured an order of magnitude more precisely than at the LHC. 

\subsection{mSUGRA}

Two issues have to be addressed to determine the fundamental parameters from the 
experimental measurements: finding the true/correct parameter set from a strongly 
correlated system of measurements and determining accurately the errors on the
parameters. 

For the first issue several different techniques have been developed to efficiently
sample a multi dimensional parameter space. Fittino~\cite{Bechtle:2005vt} 
has developed the technique of
simulated annealing, which allows to cross potential boundaries. These boundaries
could confine a simple fit based search to secondary minima. 
SFitter~\cite{Lafaye:2004cn,Lafaye:2007vs}
has developed weighted Markov chains which have the advantage of efficient sampling
in high dimensions. Markov chains are linear in the number of parameters in contrast to a simple
grid based approach where all parameters are sampled with a predefined step size.

\begin{table*}[htb]
\caption[]{Expected errors on the mSUGRA parameters at the LHC and
ILC. Flat theory errors are used.}
\begin{center}
\begin{tabular}{|l|r|ccc|ccc|}
\hline
            & SPS1a  
                     & $\Delta_{\rm endpoints}$ 
                     & $\Delta_{\rm ILC}$ 
                     & $\Delta_{\rm LHC+ILC}$ 
                     & $\Delta_{\rm endpoints}$ 
                     & $\Delta_{\rm ILC}$ 
                     & $\Delta_{\rm LHC+ILC}$ \\
\hline
            &        & \multicolumn{3}{c|}{exp. errors}
                     & \multicolumn{3}{c|}{exp. and theo. errors} \\
\hline
$m_0$       & 100    & 0.50 & 0.18  & 0.13  & 2.17 & 0.71 & 0.58 \\
$m_{1/2}$   & 250    & 0.73 & 0.14  & 0.11  & 2.64 & 0.66 & 0.59 \\
$\tan\beta$ & 10     & 0.65 & 0.14  & 0.14  & 2.45 & 0.35 & 0.34 \\
$A_0$       & -100   & 21.2 & 5.8   & 5.2   & 49.6 & 12.0 & 11.3 \\
$m_t$       & 171.4  & 0.26 & 0.12  & 0.12  & 0.97 & 0.12 & 0.12 \\
\hline
\end{tabular}
\end{center}
\label{tab:sugra_ilc}
\end{table*}

\begin{figure}[htb]
\includegraphics[width=\columnwidth]{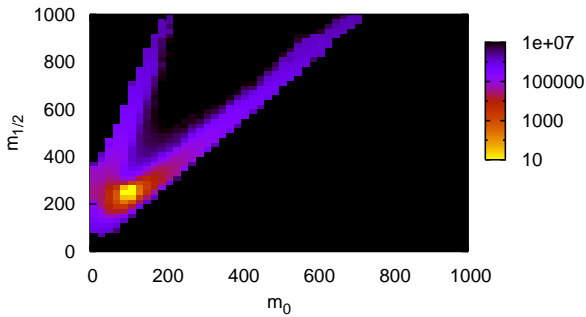}
\vspace{-1.5cm}
\caption{The result of the profile likelihood in the 
$(\mathrm{m}_{1/2}/\mathrm{m}_0)$ plane is shown~\cite{Lafaye:2007vs}. The best--fit point
is clearly identified and corresponds to the true SPS1a parameter set.}
\label{fig:m0m12mSUGRA}
\end{figure}

After having produced a full dimensional exclusive likelihood map, two types
of projections are possible to reduce the number of parameters, e.g., to illustrate
correlations between two parameters. The Bayesian approach introduces a measure
but conserves the property of probability density in the projection.
The frequentist approach uses a profile likelihood which ensures that the absolute
minimum is always retained in the projection. In illustration of a two dimensional
correlation plot is shown in Figure~\ref{fig:m0m12mSUGRA} for the frequentist approach
in mSUGRA. The correct parameter set was found from the measurements. 

The Markov chains also allow to identify secondary minima. Even 
in a tightly constrained model such as mSUGRA secondary minima occur. In particular,
when in addition to the fundamental parameters the Standard Model parameters are
also allowed to vary within their experimental error. SFitter showed~\cite{Lafaye:2007vs} 
that the interplay of the top quark mass and the tri--linear coupling can lead 
to additional solutions. However these can be discarded easily by comparing the
$\chi^2$ value of the solutions which are much larger than for the correct parameter set.

To address the second issue, determination of the errors on the parameters, one can 
either determine them in a single fit using for example MINOS or by using toy-experiments,
i.e., performing the parameter determination for a large number of datasets which have
been smeared according to the experimental and theoretical errors. In the latter 
case the width (RMS or a Gaussian fit) of the distribution of the central value of the fits
is the error. This procedure is more robust than a single fit and used most of the time,
especially in the presence of correlations.

While the experimental errors can usually be treated as Gaussian with well defined correlations,
e.g., the energy scale error is fully correlated among measurements, the theoretical
error deserves special attention. Here the RFit~\cite{Hocker:2001xe} approach is followed.
Within the theoretical error, the contribution to the $\chi^2$ is zero. Beyond this region
the usual $\chi^2$ contribution calculated with the experimental error is used. 
The procedure ensures that no particular value within the theoretical error region
is privileged. The typical theoretical errors used for this study are 3\% for strongly
interacting particles (squarks and gluinos) and 1\% for weakly electromagnetically 
interacting particles (sleptons, neutralinos and charginos). In the Higgs sector an error of 
2~GeV is used. 

The results are summarized in Table~\ref{tab:sugra_ilc} for the LHC, ILC, the combination of 
LHC and ILC with and without theoretical errors. The precision
of the LHC alone is at the level of percent for the determination of the parameters. It is improved
by the ILC by about an order of magnitude. Including the theoretical errors has an impact on the
precision at both machines, the errors are larger by a factor of three to four. Thus the precision
of the parameter determination at the LHC  
is limited by the precision of the theoretical predictions.
The SPA project~\cite{AguilarSaavedra:2005pw} aims to improve the situation.
It is also important to note that at the LHC the precision on the top quark 
mass of 1~GeV has an impact of about 10\% on the precision of the 
determination of the parameters.

\subsection{MSSM}

\begin{figure}[htb]
\includegraphics[width=\columnwidth]{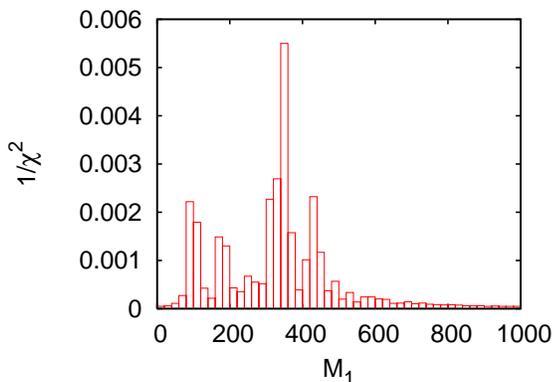}
\vspace{-1.5cm}
\caption{The result of the profile likelihood for the gaugino mass parameter 
$\mathrm{M}_{1}$ is shown~\cite{Lafaye:2007vs}. 
Several peaks are identified corresponding to secondary maxima.}
\label{fig:M1MSSM}
\end{figure}

The search for the global minimum is more complex in the MSSM where more 
parameters have to be determined. If the unification of the first and generation
sfermions is not assumed, 19~parameters have to be determined. A complex mix of different
techniques has to be applied: markov chains and a gradient search to refine the resolution
of the minima. The result of the profile likelihood procedure at the LHC is shown in 
Figure~\ref{fig:M1MSSM} for the gaugino mass parameter $\mathrm{M}_{1}$. 

\begin{figure*}[htb]
\includegraphics[width=\columnwidth]{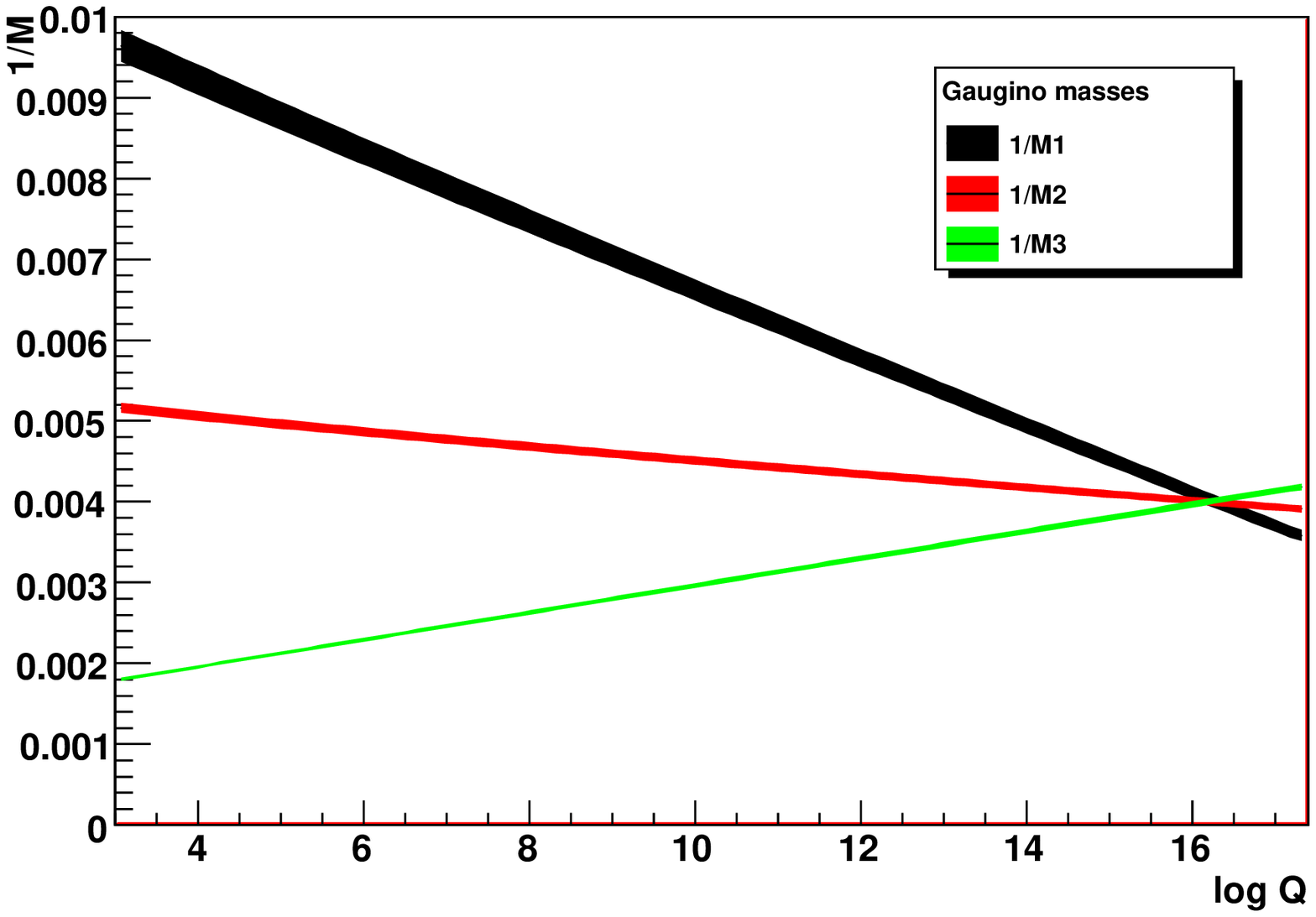}
\hspace{0.5cm}
\includegraphics[width=\columnwidth]{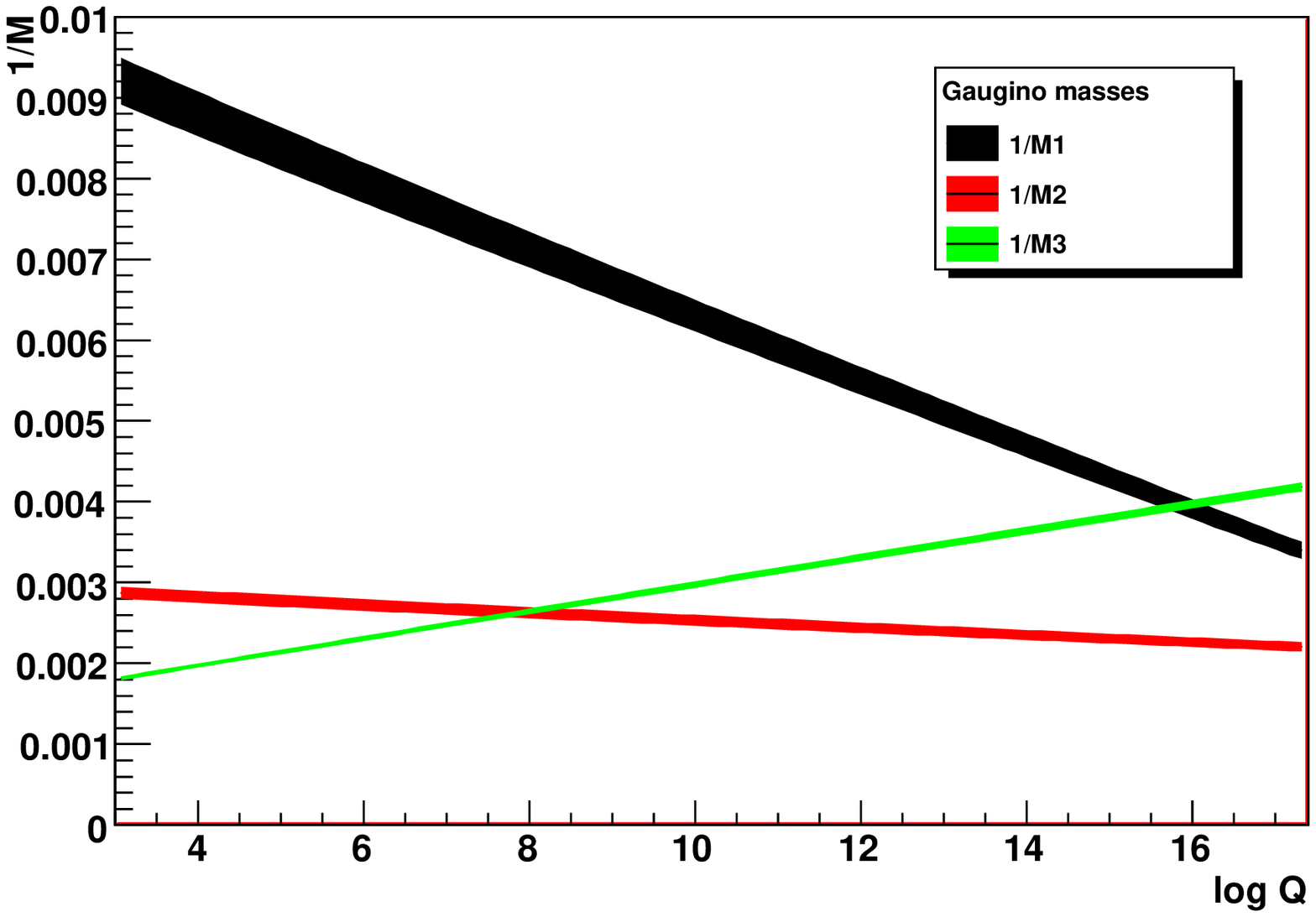}
\includegraphics[width=\columnwidth]{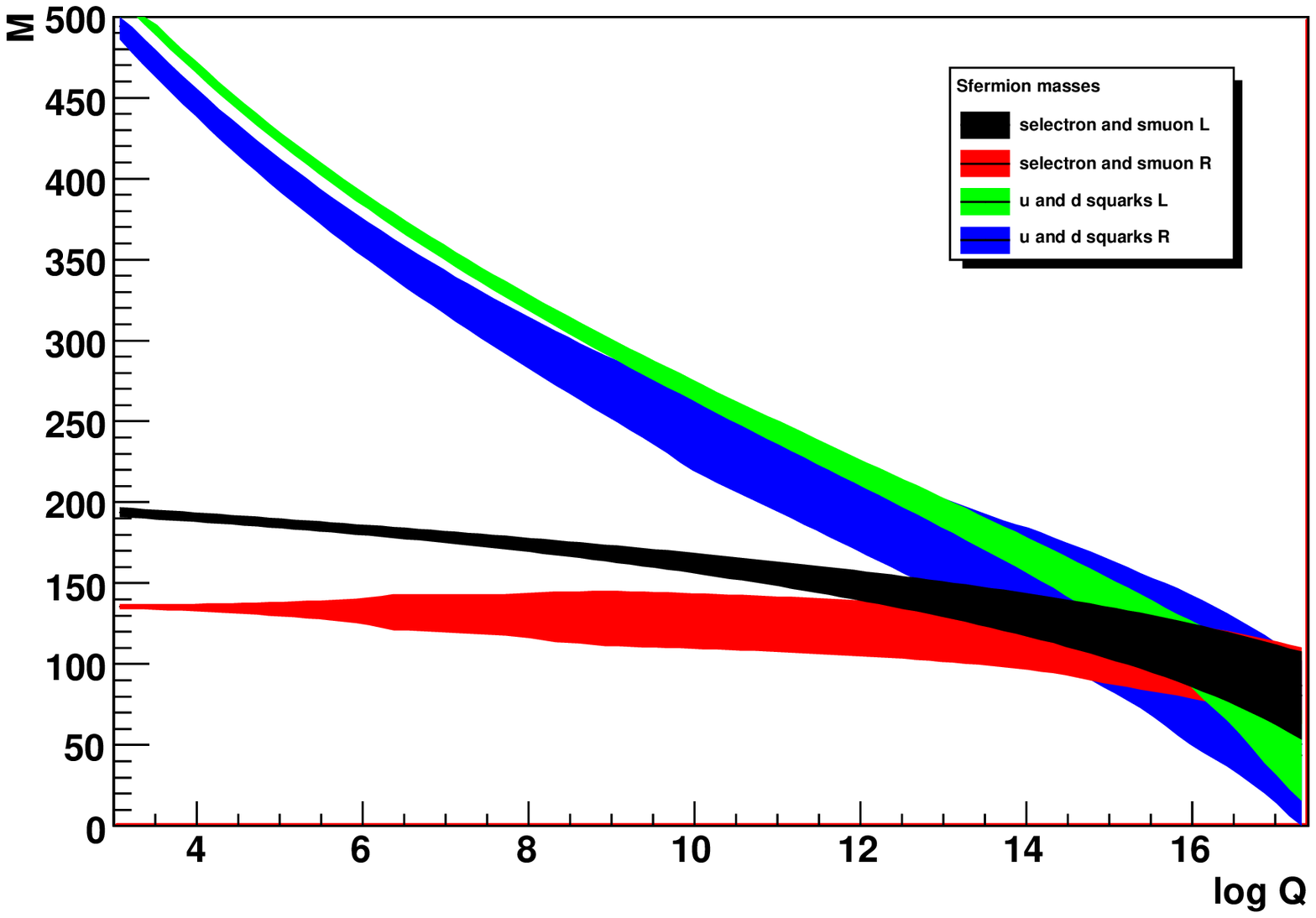}
\hspace{0.7cm}
\includegraphics[width=\columnwidth]{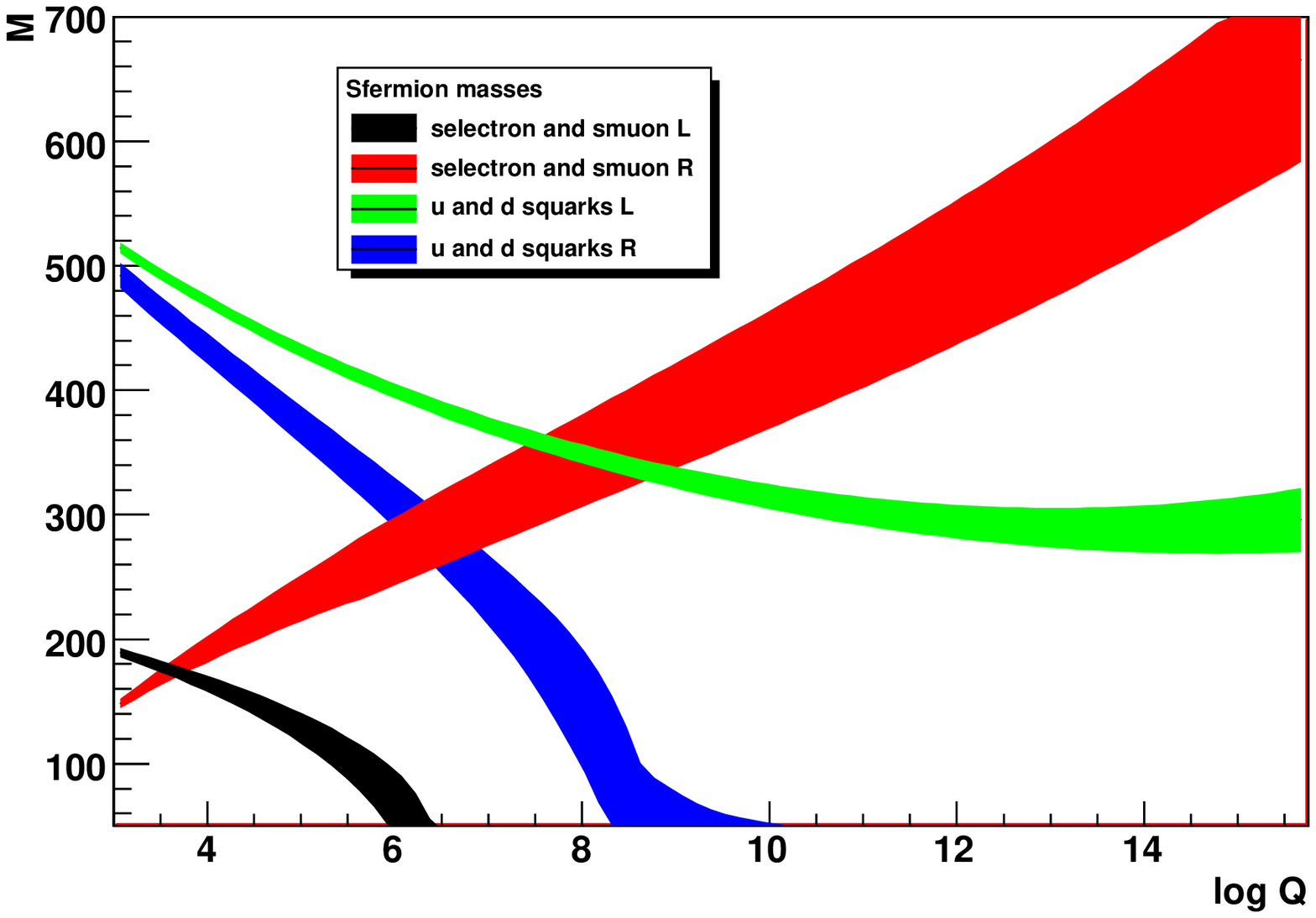}
\vspace{-0.5cm}
\caption{Top: Extrapolation of the inverse of the gaugino mass parameters to the GUT scale for 
two of the degenerate solutions at the LHC. Grand unification
is observed in one case corresponding to the correct parameter set (left), but not in the 
other solution (right). M$_1$ is in black, M$_2$ red and M$_3$ is in green.
Bottom: Extrapolation of the first and second generation scalar mass parameters to the GUT scale for 
two of the degenerate solutions. Grand unification
is observed in one case (left) 
corresponding to the correct SPS1a MSSM parameter set, but not the other one (right).}
\label{fig:SPS1aExtrapol}
\end{figure*}

Several secondary minima are observed. They cannot be distinguished by their 
$\chi^2$ value
which is identical in all cases. The degeneracy of the eight secondary 
minima can be understood qualitatively by analyzing the gaugino--Higgsino
sector. At the LHC in SPS1a only three neutralinos can be measured according
to the studies available today, but the sector is determined by four parameters
M$_1$, M$_2$, $\mu$ and $\tan\beta$, neglecting radiative corrections.
The system is therefore under-constrained. Additionally 
M$_1$ and M$_2$ can be exchanged without a change in the LHC observables.
The situation is improved dramatically when the ILC measurements are added as
they complete the neutralino sector and add the masses of the charginos 
providing additional constraints to determine the MSSM parameters 
without ambiguity. 

The MSSM is defined at the electroweak scale. The definition of its parameters 
does not depend on the model of supersymmetry breaking. Thus, having determined the
parameters at the weak scale, the extrapolation of 
the parameters to the GUT scale can be performed. 
Instead of assuming unification at the GUT scale as in mSUGRA, it can now be tested. 
As eight degenerate solutions have been determined the extrapolation 
is performed for each one of them. Two examples are shown in 
Figure~\ref{fig:SPS1aExtrapol} (top). Figure~\ref{fig:SPS1aExtrapol} (top left)
corresponds to a solution which is the correct SPS1a MSSM parameter set. GUT unification
of the gaugino masses is observed as expected. However in 
Figure~\ref{fig:SPS1aExtrapol} (top right) GUT unification is not observed. 
Of the eight degenerate solutions only one unifies the gaugino masses at the GUT scale, 
one almost does (this corresponds to the parameter set where only the sign of $\mu$ is changed 
with respect to SPS1a) and the six other solutions clearly do not unify.

In addition to the gaugino sector the extrapolation of the scalar sector in the 
first two generations can also be studied at the LHC. The third generation is under-constrained
due to the lack of measurements in the stop sector.
In Figure~\ref{fig:SPS1aExtrapol} (bottom) the extrapolation is shown on the left for 
the correct SPS1a MSSM parameter set and on the right for the same set with the exception of the 
stau parameter which has been moved far from the true solution (the $\chi^2$ values are identical).
Unification can be observed in the first case, but not in the second one, while the gaugino
unification is observed in both cases. This can be understood from the RGE equations. 
The RGEs in the gaugino sector essentially decouple, whereas in the RGEs of the scalar 
sector for a parameter all other parameters. A bad measurement, i.e., far from its true
value, therefore has a large impact on the extrapolation of all scalar parameters.
The extrapolation is stabilized by the measurements of the ILC. The combination of 
LHC and ILC therefore will allow to measure grand unification of all supersymmetric
parameters, whereas for the LHC, without further measurements, one can only observe that 
one of the ambiguous solutions is compatible with GUT unification.

\subsection{RELIC DENSITY}

\begin{table*}[htb]
\caption[]{Signatures included in the Higgs coupling 
analysis for a Higgs mass of 120~GeV and an integrated luminosity of
30~fb$^{-1}$. The factor after the background rates 
  describes how many events are used to extrapolate into the signal 
  region. The last two columns give the one-sigma experimental and 
  theory error bars on the signal.} 
\begin{center}
\begin{small}  
\begin{tabular}{l|l||r|r@{ }l|r||r|r} 
 production & decay &  
 $S+B$ &  
 \multicolumn{2}{|r|}{$B$ }&  
 $S$ &  
 $\Delta S^\mathrm{(exp)}$ &   
 $\Delta S^\mathrm{(theo)}$ \\ \hline 
 $gg \to H$ & $ZZ$ &  
  13.4 & 6.6 & ($\times$ 5) & 6.8 & 3.9 & 0.8 \\ 
 $qqH$ & $ZZ$ &  
  1.0  & 0.2 & ($\times$ 5) & 0.8 & 1.0 & 0.1 \\ 
 $gg \to H$ & $WW$ &  
  1019.5 & 882.8 & ($\times$ 1) & 136.7 & 63.4 & 18.2 \\ 
 $qqH$ & $WW$ &  
  59.4 & 37.5 & ($\times$ 1) & 21.9 & 10.2 & 1.7 \\ 
 $t\bar{t}H$ & $WW (3 \ell)$ &  
  23.9 & 21.2 & ($\times$ 1) & 2.7 & 6.8 & 0.4 \\ 
 $t\bar{t}H$ & $WW (2 \ell)$ &  
  24.0 & 19.6 & ($\times$ 1) & 4.4 & 6.7 & 0.6 \\ 
 inclusive & $\gamma\gamma$ &  
  12205.0 & 11820.0 & ($\times$ 10) & 385.0 & 164.9 & 44.5 \\ 
 $qqH$ & $\gamma\gamma$ &  
  38.7 & 26.7 & ($\times$ 10) & 12.0 & 6.5 & 0.9 \\ 
 $t\bar{t}H$ & $\gamma\gamma$ &  
  2.1 & 0.4 & ($\times$ 10) & 1.7 & 1.5 & 0.2 \\ 
 $WH$ & $\gamma\gamma$ &  
  2.4 & 0.4 & ($\times$ 10) & 2.0 & 1.6 & 0.1 \\ 
 $ZH$ & $\gamma\gamma$ &  
  1.1 & 0.7 & ($\times$ 10) & 0.4 & 1.1 & 0.1 \\ 
 $qqH$ & $\tau\tau (2 \ell)$ &  
  26.3 & 10.2 & ($\times$ 2) & 16.1 & 5.8 & 1.2 \\ 
 $qqH$ & $\tau\tau (1 \ell)$ &  
  29.6 & 11.6 & ($\times$ 2) & 18.0 & 6.6 & 1.3 \\ 
 $t\bar{t}H$ & $b\bar{b}$ &  
  244.5 & 219.0 & ($\times$ 1) & 25.5 & 31.2 & 3.6 \\ 
 $WH/ZH$ & $b\bar{b}$ &  
 228.6 & 180.0 & ($\times$ 1) & 48.6 & 20.7 & 4.0  
\end{tabular} 
\end{small} \vspace*{0mm} 
\end{center}
\label{tab:HiggsChannels} 
\end{table*} 

At this point all supersymmetric parameters have been determined, therefore the
complete particle spectrum can be deduced. From the spectrum and its couplings
the relic density is predicted~\cite{Baltz:2006fm}. Neglecting the theoretical
error and performing the analysis in mSUGRA, a precision of about 2\% is
expected at the LHC with an improvement by an order of magnitude expected
from the ILC. The precision depends on the exact nature of the parameter
set and is not valid in all scenarios. The precision compares well to the one expected 
by the Planck satellite which was successfully launched recently. Thus the confrontation
of the collider prediction of the relic density and the measurement of the 
cosmic microwave background fluctuation will provide for interesting physics 
studies in the future.

\section{DIFFICULT SUPERSYMMETRY}

While one can hope for the LHC discovery of SPS1a-like supersymmetry as hinted by
the analysis of precision observables of the electroweak sector, b-sector and the relic density, 
difficult scenarios must also be envisaged. One of these
is DSS, also known as Split-SUSY (\cite{Kilian:2004uj,Bernal:2007uv} and references therein). 
In these models the scalars are too heavy to be produced at the LHC. However
several observables are still present: the mass of the lightest Higgs boson,
the mass difference between the second lightest neutralino and the LSP,
the cross section of the tri--lepton signal, ratios of the decay of supersymmetric
particles to the Z boson and the gluino production cross section. 
Putting all of this together, the DSS parameters can be determined and measured
at the LHC. This is an indication that the LHC can also handle difficult scenarios.

\subsection{HIGGS}

\begin{figure*}[htb]
\vspace{-0.5cm}
\includegraphics[width=\columnwidth]{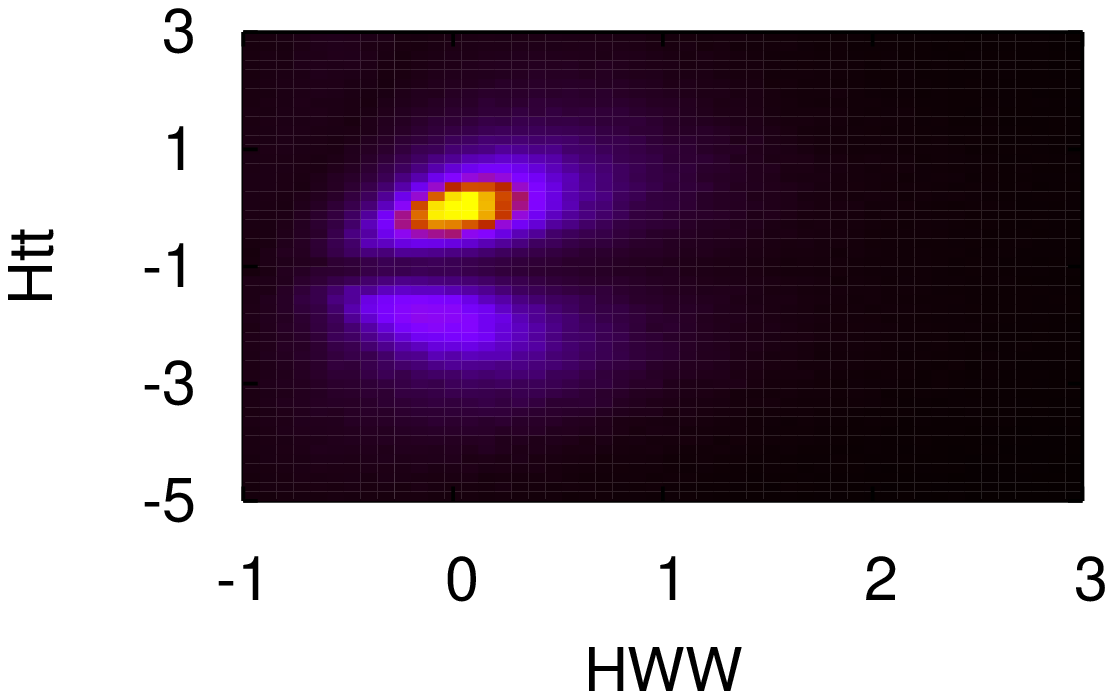}
\hspace{1cm}
\includegraphics[width=\columnwidth]{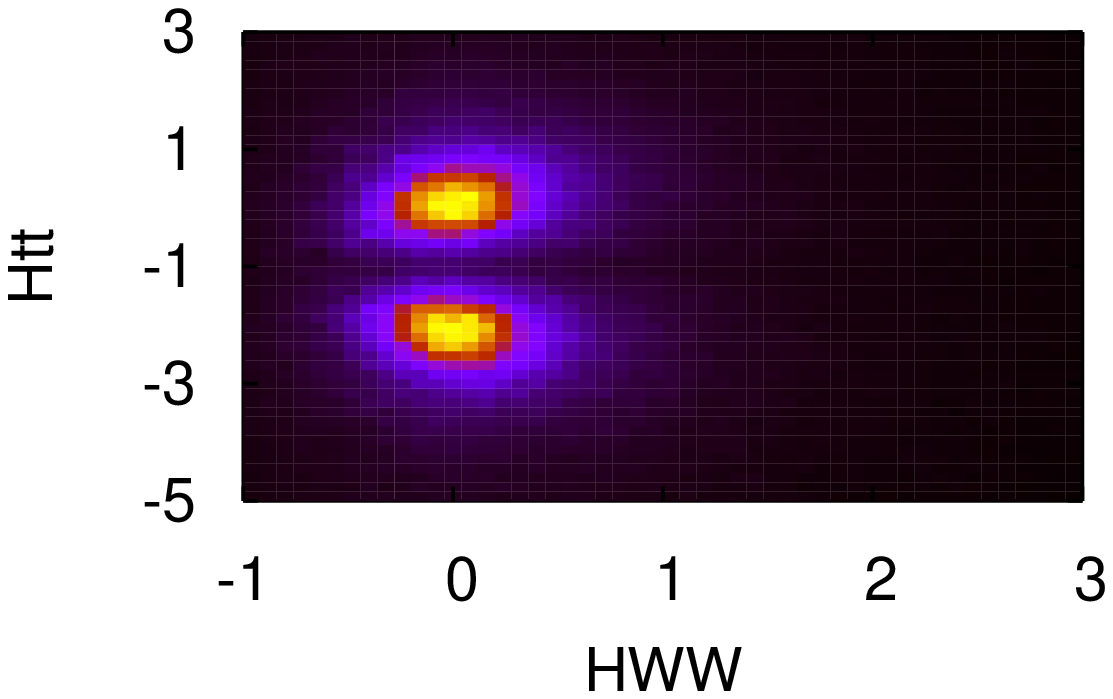}
\vspace{-1.0cm}
\caption{The result of the profile likelihood is shown for the Htt versus the HWW couplings 
on the (left) without allowing additional contributions to effective couplings and 
on the (right) allowing additional contributions to effective couplings.}
\label{fig:HiggsCouplings}
\end{figure*}

Another difficult scenario at the LHC would be that only a light boson
with a mass of about 120~GeV is discovered at the LHC but no 
(other) new particles. The expected measurement channels and 
their experimental and theoretical errors are shown in Table~\ref{tab:HiggsChannels}.
While an ILC is capable of a model independent search for the Higgs boson, at the LHC
the search and measurement is restricted to well defined final states. 
Studying the precision of the determination of the Higgs couplings could shed 
light on new physics.  

The study presented here~\cite{Lafaye:2009vr} uses essentially the same 
MonteCarloData as~\cite{Duhrssen:2004cv} with two important differences:
In agrement with more recent experimental studies by ATLAS and 
CMS~\cite{Ball:2007zza,Aad:2009wy}, 
the ttH$\rightarrow$bb signal is reduced by 50\%. On the other hand the
new theoretical analysis of the ZH($\rightarrow$bb) (subjet analysis)~\cite{Butterworth:2008iy}
is used, confirmed by ATLAS within 10\%~\cite{ATLAS:2009hw}. 

The theoretical errors on the production cross section are between 7\% and 13\%. The errors
on the branching ratio calculations are between 1\% and 4\% (c quarks). The experimental
errors have a statistical uncorrelated component and a several sources of systematic 
error which are highly correlated among measurements. Determining the Higgs couplings from
this system is therefore similar to the determination of supersymmetric parameters, i.e.,
the same techniques can be used to extract the couplings. The additional difficulty to be 
mastered as some of the channels are in the Poisson regime is the convolution 
of the flat, Gaussian and Poisson errors.

The parameters are defined in the following way for tree level couplings:
\begin{equation}
 g_{jjH} \longrightarrow g_{jjH}^{\mathrm{SM}} \; \left( 1 + \Delta_{jjH}
                                   \right)
\end{equation}
as deviations from the Standard Model value. For the loop induced 
couplings such as the $\gamma\gamma$H and ggH couplings, the definition
is as follows
\begin{equation}
 g_{jjH} \longrightarrow 
  g_{jjH}^{\mathrm{SM}} \; \left(
   1 + \Delta_{jjH}^{\mathrm{SM}} + \Delta_{jjH} \right)
\end{equation}
The additional term $\Delta_{jjH}^{\mathrm{SM}}$ is the modification of the jjH 
effective coupling induced by 
a deviation of the tree level couplings of the particles in the loop.

As in the supersymmetric case, first a full dimensional likelihood map is calculated
from which the Bayesian and frequentist projections are possible. In the Higgs 
analysis however
no true secondary 
minima exist and the noise effects of the Bayesian integration are large, essentially
washing out the signal. Therefore only the frequentist projection is pursued.

Two main observations can be made after the projections to two dimensions: there is
a general positive correlation among all couplings with the exception of the bbH coupling. 
This is due to the fact that for a Higgs boson mass of 120~GeV the total width is dominated
by the decay to b-quarks (approximately 90\%). The total width enters in the denominator
of every observable, whereas the individual partial widths, proportional to the 
couplings (squared) enter in the numerator. 

The second observation, 
as shown in Figure~\ref{fig:HiggsCouplings} (left), is that as long as no additional
contributions in the loop induced couplings are allowed ($\Delta_{jjH}$ fixed to 0),
a preference for the correct sign of the ttH coupling with respect to the WWH coupling
is observed. This is due to the Higgs decay channel to photons. Allowing genuine anomalous
couplings in the loop induced couplings destroys this distinction completely as the 
additional parameter allows for the compensation of the sign preference
without a penality to be paid in the $\chi^2$ value.

\begin{figure*}[htb]
\includegraphics[width=\columnwidth]{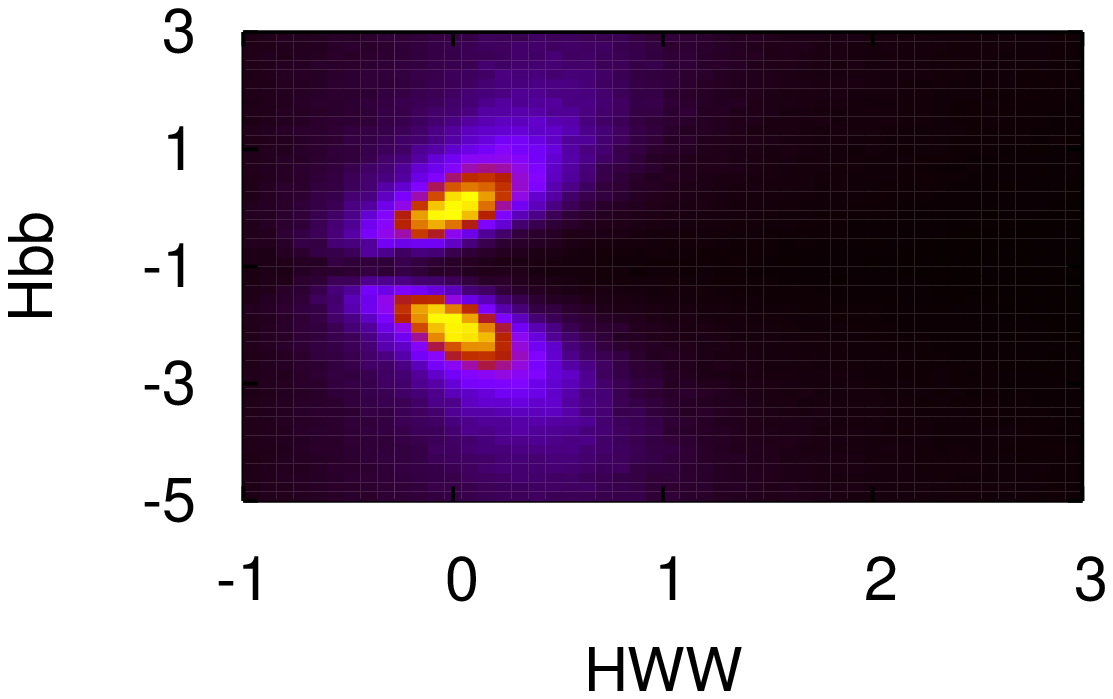}
\hspace{1cm}
\includegraphics[width=\columnwidth]{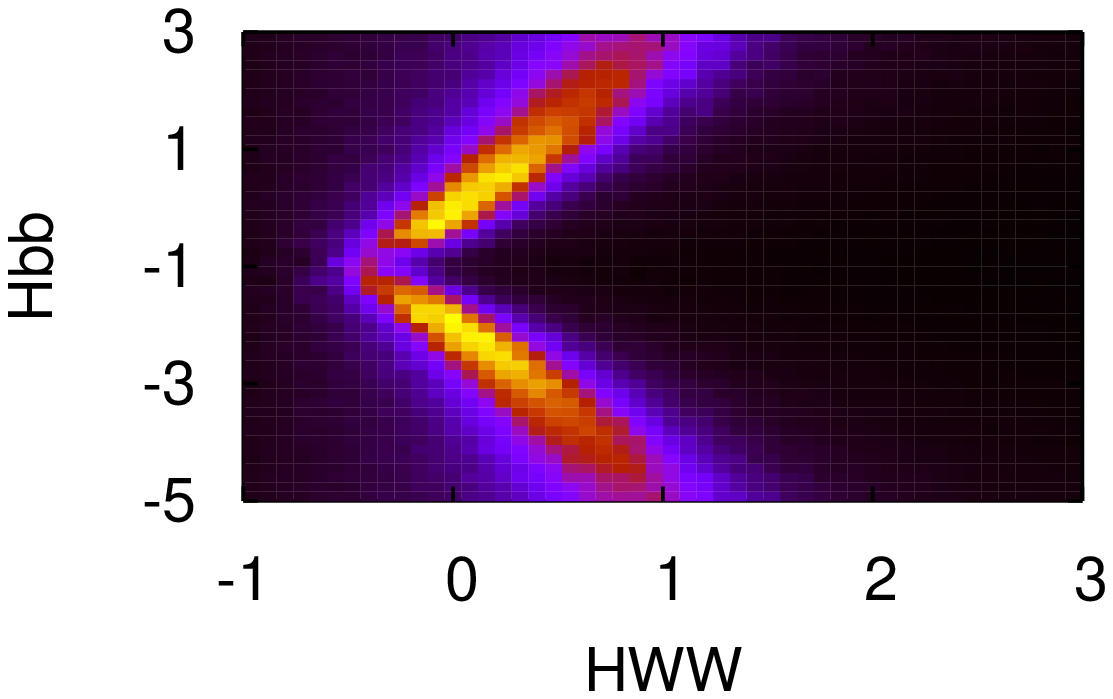}
\vspace{-1.0cm}
\caption{The impact of the subjet analysis on the Higgs boson coupling determination
is shown. On the (left) the full sensitivity is used, on the (right) the channel 
is removed completely.}
\label{fig:HiggsSubjet}
\end{figure*}

It is also illustrative to remember what the LHC cannot do in addition to what it can do.
All observables are have the structure (simplified) of cross section times branching ratio. 
Expressed in partial widths we have $\Gamma_{iiH}\Gamma_{jjH}/\Gamma_{tot}$. As partial
widths are proportional to the square of the couplings, the structure becomes
$g^2 g^2/\Gamma_{tot}$. Thus if a coupling not measured at the LHC, e.g. Hcc is much larger 
than expected, the total width will increase, but the coupling determination would 
show deviations from the Standard Model, i.e., couplings smaller than expected. 
A simultaneous determination of the individual couplings and the total width
is not possible as no model independent channel is available. Thus the LHC can measure 
absolute couplings only under the assumption that all non--measured couplings
are equal to their Standard Model prediction.

With this caveat in mind, assuming an integrated luminosity of 30~fb$^{-1}$,
the Higgs boson couplings can be determined with a precision of 
30\% (WWH) to 50\% (ttH). As before toy experiments were used to obtain
these results. The precision
is somewhat affected~\cite{Lafaye:2009vr} by whether effective couplings are
allowed to vary or not. The effect is of the order of up to 20\% (relative).
If the ratios of couplings are analysed instead of the couplings and 
the WWH coupling is used 
as reference, the precision is of the order of 30\%. While part
of the theoretical and experimental error cancel and therefore one 
could expect some improvement, with an integrated luminosity of 30~fb$^{-1}$
the statistical error is still dominant. The conclusions could therefore 
change as more data will become available.

In Figure~\ref{fig:HiggsSubjet} (left) the (bbH,WWH) coupling plane is shown 
using the full sensitivity of the analysis. In the same Figure (right)
the impact of removing the analysis completely is shown. The spread 
of the central region is much larger. The subjet analysis is essential for the 
Higgs coupling determination at the LHC because of the importance
of the bbH coupling for low Higgs boson masses.

Once the Higgs couplings are determined one can ask whether the precision 
is sufficient to exclude new physics. The task looks daunting given
the precision of the individual couplings. 
However not only the individual couplings are important but also 
their correlations. The loglikelihood is used 
as an estimator take all correlations into account.

In the gluophobic Higgs~\cite{Djouadi:1998az} model the stop quark and top quark 
contributions cancel to reduce the ggH coupling to 24\% of the Standard Model
value, i.e. the cross section for gluon fusion processes is reduced by
a factor 25. In this case at 90\% C.L. 46\% of the toy experiments 
are not described by the Standard Model.

In a second scenario the parameter set SPS1a was moved out of the decoupling region
by modifying the mass of the CP--odd Higgs boson A, the top tri--linear coupling and $\tan\beta$. 
Some of the channels are enhanced, others are reduced by almost 40\%. 
At 90\% C.L. 77\% of the toy experiments are not described by the Standard Model.

\section{CONCLUSIONS}

If vintage supersymmetry, as hinted by the electroweak fits,
is found at the LHC, a large number
of measurements will be available which can be translated into 
the fundamental parameters of mSUGRA and even the MSSM. 
To unambiguously determine the
parameter set and measure grand unification of the breaking parameters
however the ILC will be necessary.
The LHC is prepared to determine the fundamental parameters in difficult
scenarios, e.g., if only a Higgs boson is discovered at the LHC.
A determination of the Higgs couplings can shed light on new physics. 

The extraction of the fundamental parameters of any theoretical
model is a formidable task which requires a close collaboration of experimentalists 
and theorists to develop the sophisticated analyses to determine the fundamental
parameters at the LHC (and beyond at the ILC) whatever nature has in store.

\section*{ACKNOWLEDGMENTS}

It is a pleasure to thank the organizers for the 
wonderful conference. Most of the work presented in this paper
are the results of the SFitter collaboration consisting of Tilman Plehn,
Michael Rauch and R\'emi Lafaye with Michael D\"uhrssen and Claire Adam.
I am indebted to all of them for their help in the preparation
of the talk and the manuscript.
Part of the work was developed in the French GDR Terascale (CNRS).

\end{document}